# On the emergence of conductivity at SrTiO$_3$-based oxide interfaces – an in-situ study

*Merlin von Soosten,[1,2] Dennis. V. Christensen,[1] Chang-Beom Eom,[3] Thomas. S. Jespersen,[2] Yunzhong Chen,[1] Nini Pryds*[1]*

[1] Department of Energy Conversion and Storage, Technical University of Denmark, DTU Risø Campus, 4000 Roskilde, Denmark

[2] Center for Quantum Devices, Niels Bohr Institute, University of Copenhagen, 2100 Copenhagen, Denmark

[3] Department of Materials Science and Engineering, University of Wisconsin-Madison, Madison, Wisconsin 53706, United States

---

[*] nipr@dtu.dk

Heterostructures and crystal interfaces play a major role in state-of-the-art semiconductor devices and play a central role in the field of oxide electronics. In oxides the link between the microscopic properties of the interfaces and bulk properties of the resulting heterostructures challenge our fundamental understanding. Insights on the early growth stage of interfaces and its influence on resulting physical properties are scarce - typically the information is inferred from post growth characterization. Here, we report on real time measurements of the transport properties of $SrTiO_3$-based heterostructures while the crystal heterostructure is forming. Surprisingly, we detect a conducting interface already at the initial growth stage, much earlier than the well-established critical thickness limit for observing conductivity ex-situ after sample growth. We investigate how the conductivity depends on various physical processes occurring during pulsed laser depositions, including light illumination, particle bombardment by the plasma plume, interactions with the atmosphere and oxygen migration from $SrTiO_3$ to the thin films of varying compositions. Using this approach, we propose a new design tool to control the electrical properties of interfaces in real time during their formation.

Since the discovery of a two-dimensional electron gas (2DEG) at the interface between two band insulating oxides, $SrTiO_3$ (STO) and $LaAlO_3$ (LAO)[1], a wealth of intriguing properties have emerged in this seemingly simple system. In the wake of LAO/STO, numerous other STO-based heterostructures have been formed by deposition of various oxide films on STO[2–4]. A common feature is that the properties of the interfaces can be tuned dramatically in numerous ways such as by controlling oxygen vacancies during growth and post annealing[5,6], ion bombardment[7], electrostatic gate potentials[8–10], strain[11], surface adsorbates[12,13] and light exposure[14]. Pulsed laser deposition (PLD) remains the most popular deposition technique for growing STO-based heterostructures, but during this complex deposition process STO is exposed to all the aforementioned stimuli. The laser shoots on the target and produces a plasma with an intense self-emission of visible and ultraviolet light[15]. The particles in the plasma plume travel towards the STO substrate where they impact with high kinetic energies on the order of tens of eV[15,16] and produce a large and dynamically varying electrostatic surface potential[17]. As the particles emerge at the STO surface, they condense into a film that exerts stress onto STO and allows for mass transfer of e.g. oxygen ions across the interface[2]. Lastly, the entire process occurs in deposition conditions which opens up for exchange of oxygen with the atmosphere[18] as well as adsorption of species such as water on the sample[19]. Complex processes are therefore expected to happen during the early stages of the growth, which may be of significant importance for the properties of the final film. If these processes can be understood and controlled, they will provide a new handle for tuning the interface properties in real time. This highlights the importance of studying the growth process in detail, but to date, only a few studies aim to partly illuminate these processes[20,21]. The major difficulty lies in the limited number of techniques capable of monitoring and controlling the interface in real time at the early stage of the nucleation and growth. Reflection high-energy electron diffraction (RHEED) is commonly employed to monitor the film growth, to control the film thickness, and probe the crystallinity during growth in real-time[22]. It is, however, limited to structural information when growing crystalline materials. A supplementary way of monitoring the interfaces during growth is by measuring the resistivity of the interface during the growth process *in-situ*[20,21]. We expect that the combination of *in-situ* methods such as RHEED and conductivity measurements during growth will give access to a territory where the initial growth conditions can be studied and controlled in detail, leading to new and improved properties of the interface.

Here, we measure the sheet resistance of the STO-based heterointerfaces and patterned devices in real time continuously from the early stage of the deposition until the final heterostructure is produced. The measured interface conductivity demonstrates the possibility to modulate the charge carriers at the interfaces in real time by engineering the top film and the oxygen content with an instant feedback on the properties.

**a)**

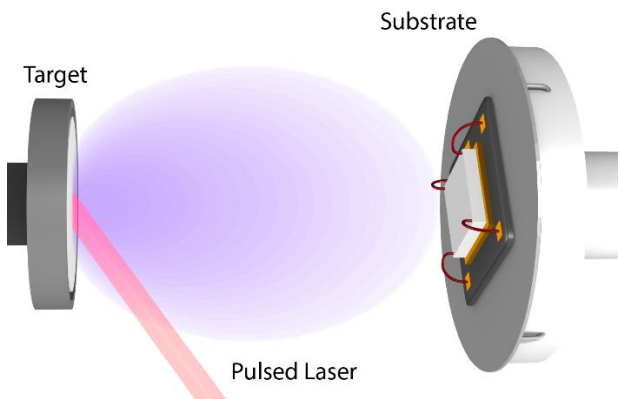

**b)**

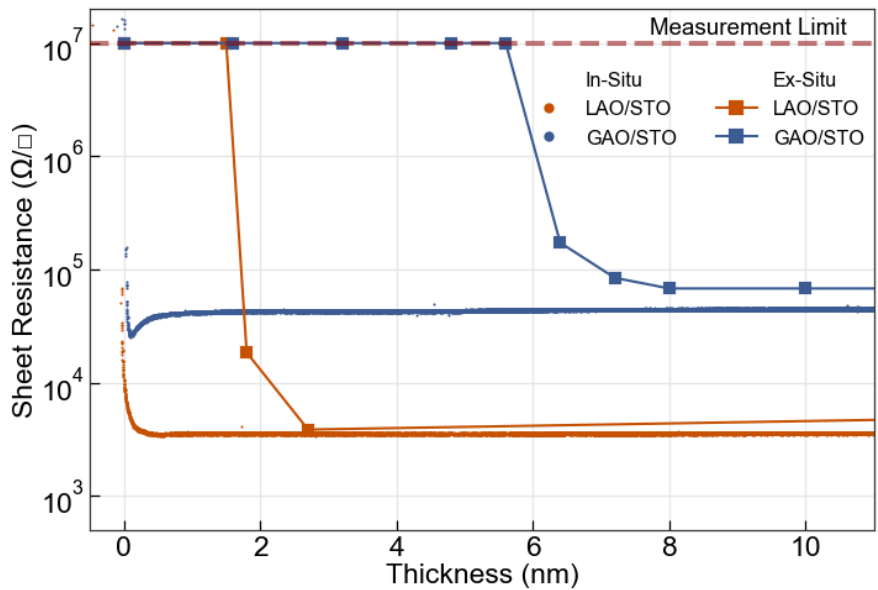

**Figure 1.** In-situ measurements of the interfacial sheet resistance during the deposition. a) Schematic illustration of the *in-situ* transport measurement system in the PLD chamber. b) Interface sheet resistance as a function of thickness for LAO/STO and GAO/STO measured *in-situ* and *ex-situ*. The *ex-situ* measurements were taken from [23] with comparable deposition conditions, we note that the critical thickness, especially for GAO, does change with conditions in [23].

*In-situ* transport measurements were carried out inside the PLD chamber during room temperature film growth on $TiO_2$ terminated STO (001) single crystals (See Fig. 1a). The evolution of the sheet resistance R$\blacksquare$ as a function of film thickness for LAO and $\gamma$-$Al_2O_3$ (GAO) deposited on STO is shown in Fig. 1b. The samples are initially insulating with R$\blacksquare > 10^7$ $\Omega/\blacksquare$ (measurement limit), but after only a few laser pulses (< 3 Pulses) the sheet resistance drops to ~$5\times10^3$ $\Omega/\blacksquare$ and ~$5\times10^4$ $\Omega/\blacksquare$ for the LAO and the GAO top layers, respectively. As the deposition is continued to a top-film thickness of 2nm, the sheet resistance reaches a steady state value of $2\times10^3$ $\Omega/\blacksquare$ and $10^5$ $\Omega/\blacksquare$ for LAO and GAO top layers, close to the *ex-situ* measurements in similar samples[2,23] (see also Fig. 1b for a comparison). Interestingly, at the initial growth stage (<2nm thickness), the measured values of the interface sheet resistance are much lower than the *ex-situ* measurements.

A different behavior was found when $LaSr_{1/8}Mn_{7/8}O_3$ (LSM) was deposited on STO as shown in Fig. 2. LSM/STO interfaces are well-known to be insulating as confirmed also in our *ex-situ* measurements after retrieving the samples from the PLD chamber. Surprisingly, however, the *in-situ* measurements reveal that the LSM films initially create conducting interfaces (< 3 pulses) which turn insulating again as the growth progresses. Deposition of GAO was also performed on insulating MgO and yttria-stabilized zirconia (YSZ), and as expected, no conductivity was measured using these substrates (Fig. 2). This confirms that the charged species in the plasma are not contributing to the measured conductivity.

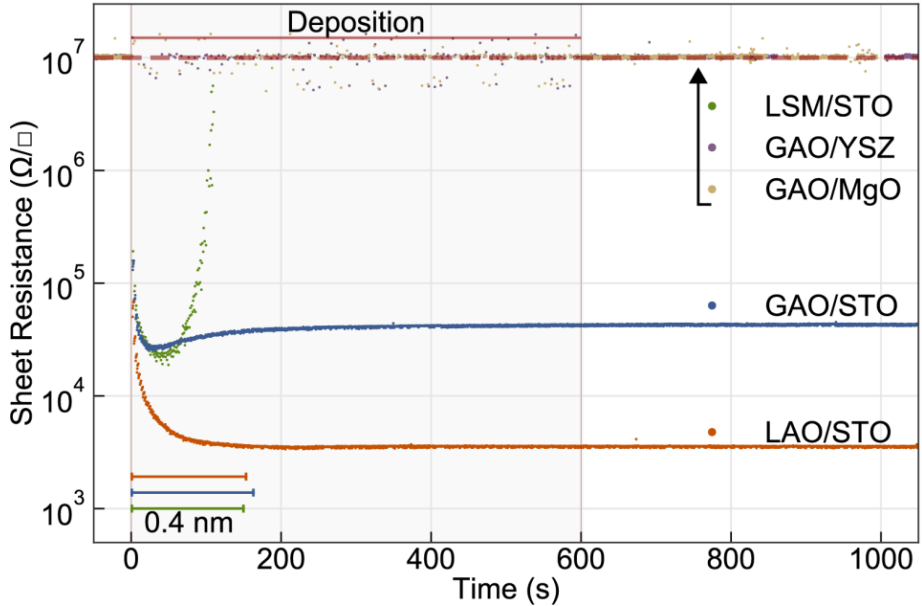

**Figure 2.** In-situ sheet resistance measurements. The measurements shown here are for the GAO/STO, LAO/STO, LSM/STO, GAO/MgO, GAO/YSZ heterostructures. Measurements on MgO and YSZ substrates are offset for clarity and remain at the measurement limit. The bar in the figure show the conversion of time to thickness for LAO, GAO and LSM.

STO and STO-based heterostructures have been reported to turn conductive when exposed to UV radiation[20,21,24–26]. To distinguish the light induced conductivity from other possible effects that can cause conductivity, we place a double-sided polished sapphire window in close proximity in front of the sample. The sapphire blocks only material from reaching the STO surface but not the UV or visible light. When the sapphire window is inserted, the deposition of LAO, GAO, and LSM show no detectable effects on treated STO substrates, and the sheet resistance remains above the measurement limit (see Fig. 3a). Also conducting samples with previously grown GAO or LAO films show no substantial effect during deposition with a sapphire window. Only samples with a high sheet resistance of $10^6$ $\Omega/\blacksquare$, show a small response to the irradiation (see Fig. S1). These samples were grown at room temperature, and subsequently annealed in air at 150C [5] in order to increase the sheet resistance to $10^6$ $\Omega/\blacksquare$.

a)

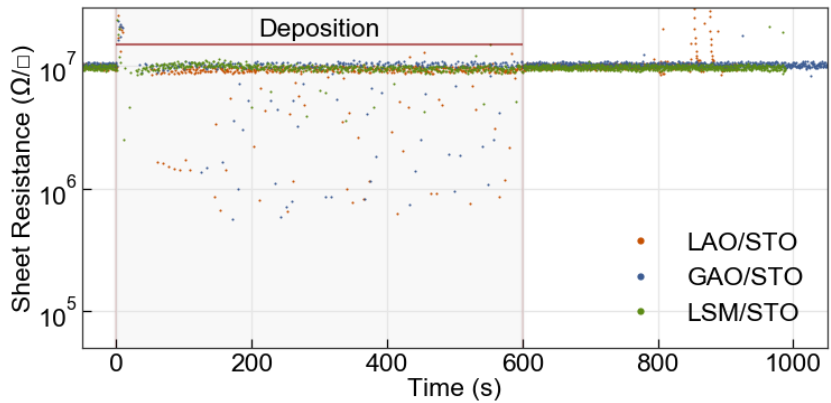

b)

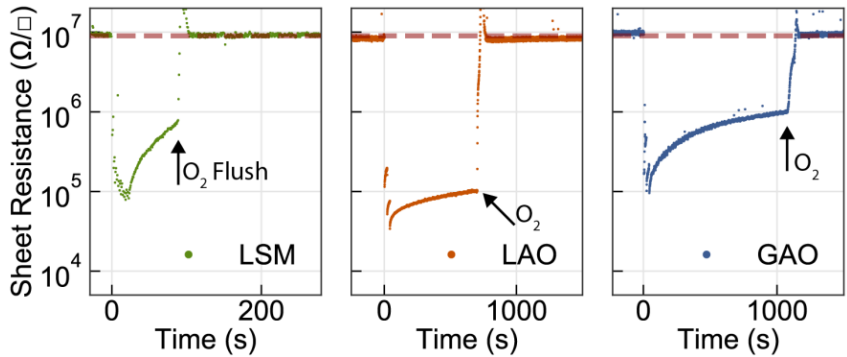

**Figure 3. a)** No measurable conductivity change is found for Sapphire covered substrates, the conductivity remains at the measurement limit throughout the deposition time. **b)**

Annealing experiments illustrating the effect of oxygen on the sheet resistance. After 3 initial laser pulses samples turn conducting and decay in 2x10$^{-6}$ mbar of Oxygen. The arrow in the figures show the onset of oxygen flushing to 5x10$^{-2}$ mbar into the chamber, and the samples turn insulating rapidly.

Several conducting STO-based interfaces have been grown at room temperature where the origin of the conductivity is attributed to oxygen vacancies[2,27,28]. In order to understand the effect of oxygen on the interfacial conductivity we carried out *in-situ* studies of the effect of background gas by flushing the PLD chamber with oxygen after the deposition, see figure 3b. Initially the laser was pulsed 3 times, and the resistance drops by several orders of magnitude, from $10^7$ to $10^4$ Ω/■ . After the short deposition, the laser was turned off and the conductivity decays in a background pressure of 2x10$^{-6}$ mbar. Using an exponential fit to the conductance (see Fig.S2) we obtain decay rates of 0.005 s$^{-1}$ (LAO), 0.006 s$^{-1}$ (GAO), and 0.081 s$^{-1}$ (LSM). Figure 3b also shows a sharp increase in sheet resistance to the detection limit, when flushing the chamber with oxygen to a pressure of 5x10$^{-2}$ mbar. Furthermore, this modulation is repeatable (see Fig. S2).

We now consider some of the mechanisms which can influence the interface conductivity and explain our experimental findings. These mechanisms are bombardment[15], light irradiation[14,21], redox reaction[2], and oxidation[5]:

*Bombardment:* During the deposition, substrates are bombarded with high energy species from the target with kinetic energies on the order of tens of eV[15], allowing loosely bound elements (i.e. oxygen) to escape from the topmost part of the substrate. This may form a conducting layer due to the formation of oxygen vacancies. During film growth, the substrate is increasingly protected by the deposited material, preventing the oxygen from leaving the interface, thus the bombardment effect is only relevant during the initial growth.

*Light Irradiation:* STO absorbs light at 3.2eV and during deposition, free charges may be generated and contribute to the conductivity of the sample.

*Redox reaction:* During growth in a low oxygen pressure environment, an oxygen deficient film is formed, and for films with a high oxygen affinity a redox reaction may take place in which oxygen is transferred from the interface region of STO to the oxygen deficient top film. This results in the formation of oxygen vacancies as well as conductivity in STO. Similar to the bombardment, the redox reaction will only take place as long as there is enough energy and a pathway for oxygen to leave the substrate. This is dependent on the oxygen affinity of the top-film and the kinetics of the oxygen transfer.

In our current work we exclude any major contributions of light induced conductivity during the depositions as confirmed from our experiments using the sapphire window. This in contrast to previous reports[20]. Regardless of the film deposited on top of STO, all interfaces studied in the current work show a dramatic drop in resistance during the first laser pulses (See Fig. 2). Even LSM deposited on STO, which is reported to highly suppress the redox reaction and create non-conducting interfaces[2], becomes conducting initially but decays very fast to its original high resistive state. We thus believe that the measured conductivity after just 3 laser pulses cannot be attributed to the redox reaction only, but this initial change in conductivity is initiated by the bombardment with the plasma species. However, these observations cannot alone explain the full behavior of the LAO, GAO, and LSM interfaces, and it appears that two additional, competing mechanisms are highly influential for determining the final conductivity of the interface.

The observed decay in the conductivity with time (see Fig. 3b), in particularly in the high oxygen pressure, suggests that oxygen available in the environment annihilates oxygen vacancies. This is mechanism supported by varying the film thickness: With thicker films the interface is protected better, resulting in lower decay rates for GAO and LAO consistent with a lower oxygen diffusion rates from the chamber to the interface.

The much higher decay rate after the small amount of LSM deposition, however, suggests that the redox reaction plays an important role. Since the top film after 3 laser shots is estimated to cover only around 15% (based on the growth rate), and hence the direct oxidation from the 85% exposed surface should be comparable for the LAO, GAO and LSM deposition. However, taking into account the redox activity of the top film may explain this difference, as LSM can itself accommodate oxygen vacancies by changing the valence of Mn. In the scenario of interface redox reaction, besides the oxygen source from the target and the background oxygen inside the chamber, oxygen ions in STO substrates also diffuse outward to oxidize the reactive plasma species absorbed on the STO surface[2]. This is consistent with previous studies where Al was found to have a much larger reducing effect on STO compared to Mn[29].

In all depositions, we initially see the same effect: A sharp drop in resistance after only a few pulses due to oxygen vacancies created by bombardment. However, as the deposition progresses different top-films show different behavior, and we explain the measurements by the following:

**LAO/STO**: The LAO film grown at room temperature has a high oxygen affinity, and Oxygen is diffusing from STO to the amorphous film. The redox reaction is dominating throughout the whole deposition process and the sheet resistance remains at a low level.

**GAO/STO**: The GAO film grown at room temperature is crystalline and allows for a lower degree of oxygen transfer across the interface compared to LAO. The deposition of GAO does not stabilize as many oxygen vacancies in STO as LAO, and with increased material deposition, oxygen diffusion is inhibited. Thus due to the competing mechanisms, the resistance increases after the initial low resistive state and stabilizes at a higher value.

**LSM/STO**: This type of top layer is different from the other two and right after the initial stage were bombardment creates oxygen vacancies, the STO surface oxidizes due to the low oxygen affinity of LSM as well as interactions with the background gas. The sheet resistance recovers therefore quickly to its initial high resistive state.

Based on the above results we were able to engineer the electronic properties of the interfaces at the atomic level by combining different sequences of materials deposited on the substrate. In Fig. 4 we show an example of the interface tunability by the combination of a sequence of deposited films such as LAO/GAO/STO. Here we combine the high quality epitaxial interface of GAO/STO with an over layer of LAO which results in a higher conductivity (compared to GAO/STO alone) due to the high carrier density caused by LAO. By varying the amount of GAO deposited on the initial layer, we control the oxygen transfer to the LAO layer which gives a fine control on the final carrier density. This provides a simple and powerful tool to not only select desired properties arising from different materials, but also to fine-tune these properties due to the immediate feedback at the initial growth stage.

So far all of these experiments were done using a van der Pauw configuration (see method section). However, in order to confirm the robustness of the *in-situ* measurements, the experiments were repeated on UV-lithography patterned Hall-bar devices[30], and no major differences were observed (see Fig. 4b).

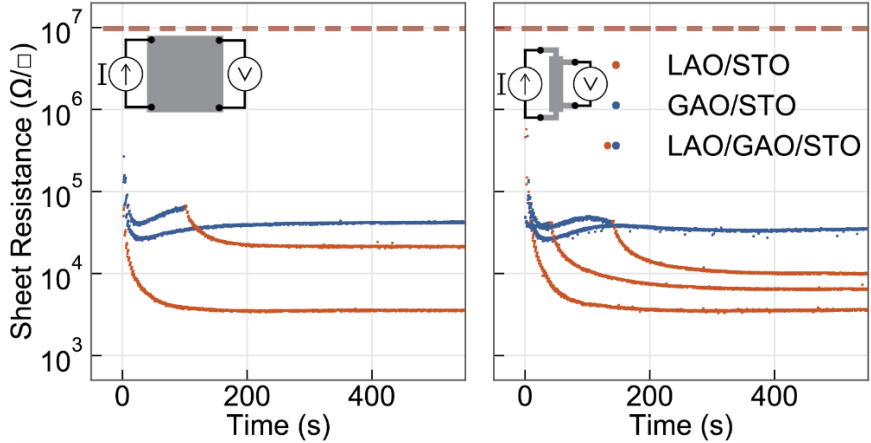

**Figure 4.** Tuning the sheet resistance of the interface via multilayers during the initial growth on a) 5x5mm samples and b) patterned Hall-Bar (0.1x0.6mm) samples. The sheet resistance in both geometries is comparable, as is the general behavior of multilayer depositions of LAO/GAO/STO.

In summary, we have shown that with a simple setup it is possible to monitor and study the sheet resistance of the interface in real-time during the early growth stage which is often not accessible during the growth process. Surprisingly, our results indicate that the conductive interface is created after the first few laser shots regardless the type of top film, in contrast to what can be deduced from *ex-situ* measurements after growth. The resulting interface conductivity is formed by a balance of bombardment, oxidation and redox reaction processes. This approach provides a new and yet undiscovered tool to tune the sample properties with real time feedback on the transport properties during the deposition by engineering the sequence of the materials deposited or by varying the deposition conditions. This provides new opportunities to design interfaces in STO based heterostructures with controlled properties.

## Methods

The interfaces were fabricated using 5 x 5 x 0.5 mm$^3$ TiO$_2$ -terminated SrTiO$_3$ (001) single crystal substrates. Amorphous LaAlO$_3$, amorphous LaSr$_{1/8}$Mn$_{7/8}$O$_3$ and crystalline γ-Al$_2$O$_3$ layers were grown by PLD at an oxygen pressure of 2x10$^{-6}$ mbar at room temperature. The thin films were grown by PLD using a KrF laser (λ = 248 nm) with a repetition rate of 0.5 Hz, and a laser fluence of 2.5 mJ cm$^{-2}$. The target–substrate distance was kept constant at 40 mm. The film thickness and crystallinity was determined by RHEED oscillations (γ-Al$_2$O$_3$, see Fig. S3) and AFM measurements (LaAlO$_3$, LaSr$_{1/8}$Mn$_{7/8}$O$_3$ and γ-Al$_2$O$_3$). The electrical resistance of the interfaces was measured during the deposition process by means of a sample carrier constructed and placed inside the PLD chamber. The sample was electrically contacted by ultrasonic wire bonding with aluminum wire. Measurements were performed using a 4-probe method in the Van der Pauw geometry. Hall-bar samples were prepared by UV-lithography, and PLD deposition was performed on the exposed and developed patterns.[30] A double-side polished sapphire plate was also placed in front of the samples, so that the sample could be shielded from the ablated particles.

## Author Contributions

M.V.S, D.V.C. and N.P. designed this project. M.V.S, D.V.C. deposited the thin films by PLD and performed the electrical characterization. M.V.S, D.V.C. and N.P. analysed the data and wrote the manuscript. M.V.S, D.V.C., C.B.E, T.J., Y.Z.C, and N.P extensively discussed the data, results, and the manuscript.

## Acknowledgements

NP would like to thank the BioWings project which has received funding from the European Union's Horizon 2020 under the Future and Emerging Technologies (FET) programme with a grant agreement No 801267. NP and DVC would like to thank also the support from the Independent Research Fund Denmark, Grant No. 6111-00145B. TSJ acknowledges funding from the Villum Foundation, Young Investigator Programme. The authors would like to acknowledge technical assistance by Jørgen Geyti.

## Competing Interests

The authors declare no competing interests.

# Supporting Information

On the emergence of conductivity at SrTiO3-based oxide interfaces – an in-situ study

*Merlin von Soosten,[1,2] Dennis. V. Christensen,[1] Chang-Beom Eom,[3]*

*Thomas. S. Jespersen,[2] Yunzhong Chen,[1] Nini Pryds[1]*

[1] Department of Energy Conversion and Storage, Technical University of Denmark, DTU Risø Campus, 4000 Roskilde, Denmark

[2] Center for Quantum Devices, Niels Bohr Institute, University of Copenhagen, 2100 Copenhagen, Denmark

[3] Department of Materials Science and Engineering, University of Wisconsin-Madison, Madison, Wisconsin 53706, United States

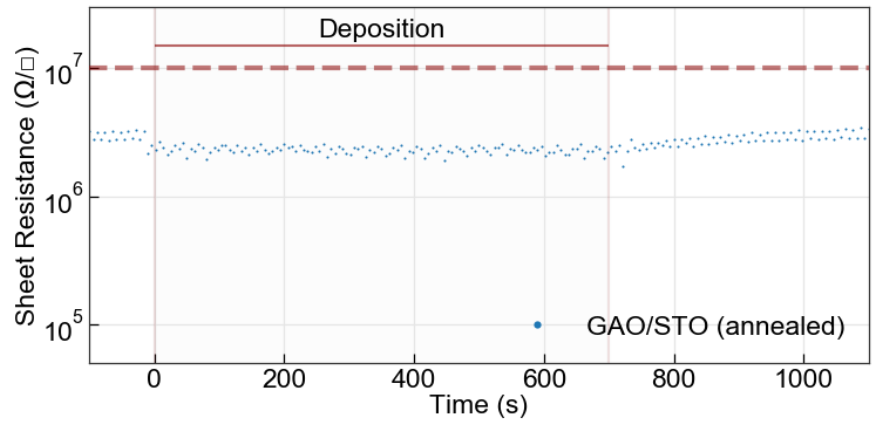

**Figure S1.** Deposition of GAO on an annealed sample through sapphire. The sample was previously grown with a GAO top film and subsequently annealed in air at 150C to reduce the sheet resistance to a stable high resistive state.

**a)**

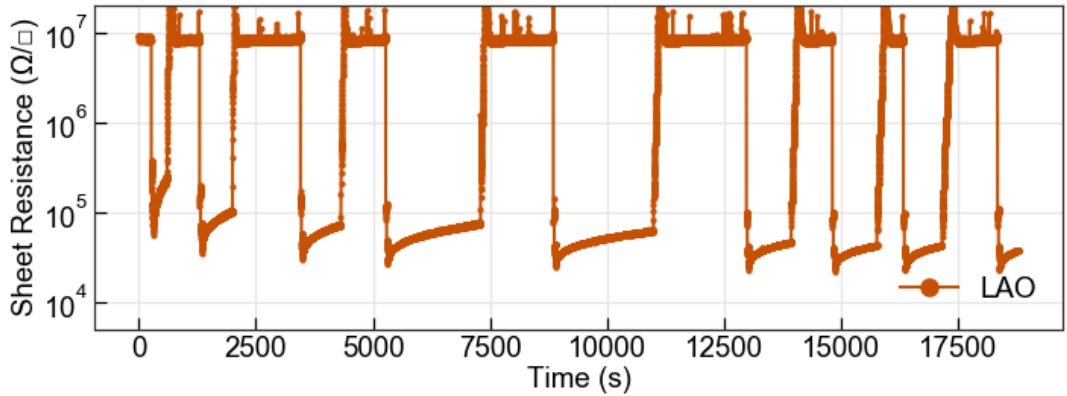

**b)**

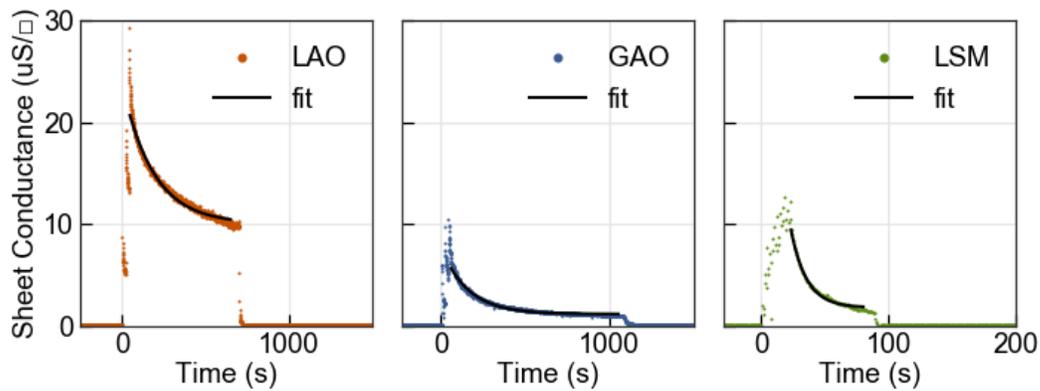

**Figure S2.** a) Oxygen annealing experiments, with sequences of: 3 laser pulses (red arrows), decay in 6e-2 mbar of oxygen background, flushing of the pld chamber with oxygen (blue arrows). Over a time of 5h 18 laser pulses were applied with an equivalent LAO thickness of 1unit cell or ~0.4nm. b) The sheet conductance decays differently for LAO, GAO, and LSM. Exponential fits to the decay reveal decay rates of 0.005 s⁻¹ (LAO), 0.006s⁻¹ (GAO), and 0.081s⁻¹ (LSM).

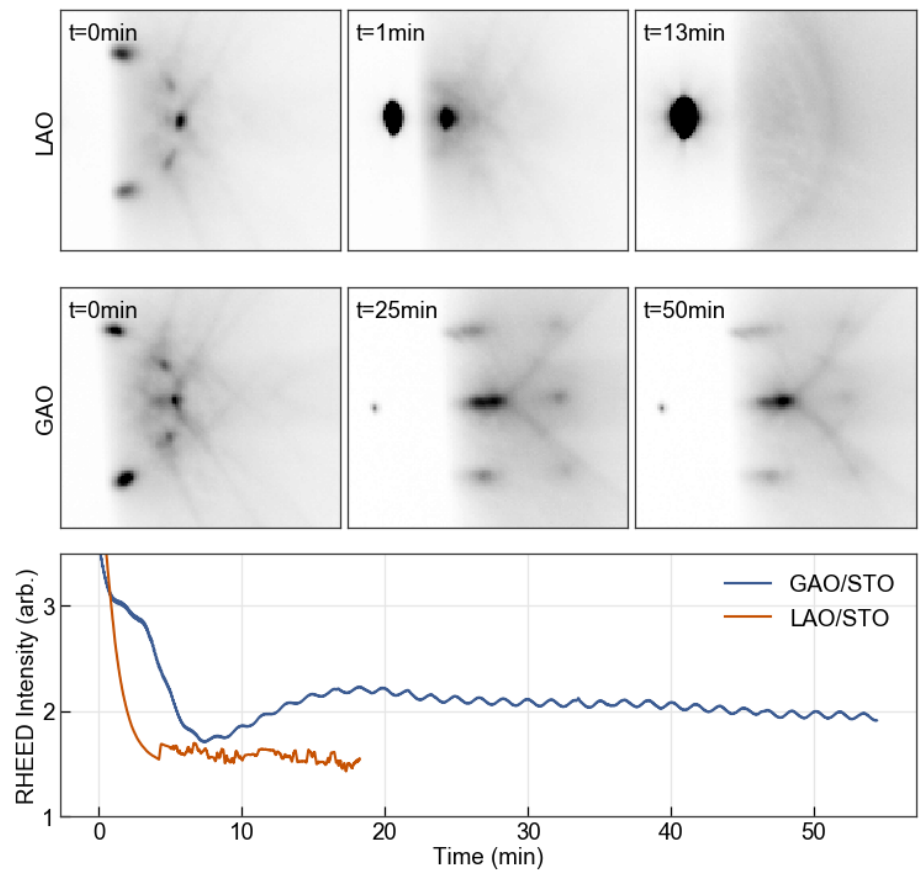

**Figure S3.** RHEED analysis of room temperature grown LAO and GAO films. The signal for LAO vanishes quickly, depicting an amorphous film growth. The GAO signal however oscillates continuously throughout the deposition time, depicting a crystalline film growth.